\documentclass[twocolumn,showpacs,prl]{revtex4}
\usepackage{graphicx}
\usepackage{dcolumn}
\usepackage{bm}

\begin{document}

\title{Delocalization of relativistic
Dirac particles  in disordered one-dimensional  systems
and its implementation with cold atoms
}
\author{Shi-Liang Zhu$^{1,2}$, Dan-Wei Zhang$^1$, and Z. D. Wang$^2$}
\affiliation{ $^1$Laboratory of Quantum Information Technology,
SPTE, South China Normal University, Guangzhou, China\\$^2$
Department of Physics and Center of Theoretical and Computational
Physics, The University of Hong Kong, Pokfulam Road, Hong Kong,
China }

\begin{abstract}
 We study theoretically  the localization
 of relativistic particles in disordered one-dimensional chains. It is found that the
relativistic particles tend to delocalization in comparison with the
non-relativistic particles with the same disorder strength. More
intriguingly, we reveal that the massless Dirac particles are
entirely delocalized for any energy due to the inherent chiral
symmetry, leading to a well-known result that particles are always
localized in one-dimensional systems for arbitrary weak disorders to
break down. Furthermore, we propose a feasible scheme to
detect the delocalization feature of the Dirac
particles with cold atoms in a light-induced gauge field.
\end{abstract}
\maketitle

Since the pioneering work by Anderson\cite{P.W.Anderson},
substantial efforts have been devoted to the understanding of
transport properties of electrons in disordered systems\cite{Lee}.
A very significant advance along this direction is the scaling
theory proposed by Thouless {\sl et al}~\cite{Thouless} and
Abrahams {\sl et al}~\cite{Abrahams}. In the scaling theory,
it is
argued that the quantity $\beta\equiv d\ln g/d\ln L$ is a monotonic
and nonsingular function of $g$ only, with $g$ a dimensionless
conductance and $L$ the sample size.  Currently there exists a
well-known result that the conductance $g$ approaches to zero as the
sample size $L$ goes to infinite for any disordered one
dimensional systems. In particular, an arbitrary weak disorder
strength leads to the localization of all states of electrons in
one-dimensional chains~\cite{Anderson1980,Abrahams,Economou}.

Actually, an implicit pre-condition for the above results is that
the particles are governed by the Schrodinger equation since the
(quasi-)particles addressed in condensed matter systems are in
general non-relativistic. Notably,  relativistic Dirac particles
have recently been paid a significant attention because the
quasi-particles in honeycomb lattices (such as electrons in the
graphene\cite{Semenoff,Novoselov} and cold atoms in the optical
lattices\cite{Zhu2007,Satija}), ultra-cold atoms in a
light-induced
gauge field\cite{Ruseckas,Juzeliunas,Stanescu,Vaishnav,Zhu2006} and
trapped ions \cite{Lamata} may be described by the relativistic
Dirac equation. However, Anderson localization in the relativistic
region has been less studied in literature\cite{Roy}. Apart from the
Klein's finding\cite{Klein} that the transmission of Dirac particles
is essentially different from that of non-relativistic particles, it
is also fundamentally important and interesting to study the
aforementioned localization issue of relativistic particles and
 to work out how to simulate the predicted results with
currently available techniques.

In this paper we study Anderson localization of relativistic
particles in disordered one-dimension systems by using the finite
scaling method and the transfer-matrix technique. 
The finite-size scaling analysis reveals that all the states of the
massive relativistic Dirac particles are localized in the systems;
while, the localization length of relativistic particles is
longer than that of
non-relativistic particles with the same disorder strength. More
intriguingly, the states of the massless Dirac particles are
entirely delocalized for arbitrary disorder strength due to the
chiral symmetry, providing a distinct example that breaks down the
well-known result for non-relativistic particles that all states
are localized in disordered one-dimensional systems. Furthermore,
we propose how to simulate the wanted Dirac particles in
disordered  one-dimensional chains and the related
(de)localization properties with recently/newly developed
techniques in the cold atomic
systems~\cite{Roati,Billy,Fallani,Lye,Lin,Chang}.


Let us consider a particle transmitting through a one-dimensional
chain with $N$ rectangular barriers as shown in Fig.1(a), where the
potential
\begin{equation}
\label{V_x} V(x)=\left\{ \begin{array}{l}
  V_n \ \ \ \ x_n \leq x \leq x_n^\prime \\
  0\ \ \ \ \ \ \text{others}
\end{array} \right.,\ \ (n=1,2,\cdots,N)
\end{equation}
with $V_n$ being a constant randomly distributed in the range
$[-\delta,\delta]$. Here $\delta$ represents the disorder
strength. For simplicity, we assume that $x_n^\prime-x_n\equiv a$
and $x_{n+1}-x_n^\prime\equiv d$ for any $n$.
A relativistic particle with the mass $m$ and energy $E$ is
inserted into the $N$ barriers from the left, which
is described by the Dirac equation
\begin{equation}
\label{Hamiltonian} [-i\hbar c\sigma_x\frac{d}{d
x}+mc^2\sigma_z+V(x)-E]\psi(x)=0,
\end{equation}
where $c$ denotes the velocity of light, $\sigma_{x,z}$ are the
Pauli matrices, and $\psi (x)$ represents a two-component spinor.
A general solution of Eq. (\ref{Hamiltonian}) for any region with
a constant potential, e.g., $V_n$ in Fig.1(a), is given by
\begin{equation}
\label{Solution}
\psi(x)=A\left(\begin{array}{c}1\\
\kappa_n\end{array}\right)e^{\frac{i}{\hbar}p_n x}+
B\left(\begin{array}{c}1\\
-\kappa_n\end{array}\right)e^{-\frac{i}{\hbar}p_n x},
\end{equation}
where $p_n$ represents the momentum of the particle,
$\kappa_n=(E-mc^2-V_n)/(cp_n)$ and $(E-V_n)^2=m^2c^4+c^2p^2_n$. If
$E$ and $V_n$ are fixed then $p_n$ can in principle be either
positive or negative.
Here we choose the positive one, and thus the coefficients $A$ and
$B$
 denote the
amplitudes of the spinors moving along the positive x-axis and its
opposite direction, respectively.

We now look into the transmission for $N$ potentials as shown in
Fig.1(a). Denoting the amplitudes of the spinor at a position
approaching to $x_n$ $(x_n^\prime)$ with the infinitesmall amount
from the left (right) as $\{ A_n,B_n\}$ ($\{
A_n^\prime,B_n^\prime\}$), we may obtain a relation between the
amplitudes based on the continuity of the wave-function,
\begin{equation}
\label{M_D}
\left(\begin{array}{c}A_{n}^\prime\\
B_{n}^\prime\end{array}\right)=M_n^D\left(\begin{array}{c}A_{n}\\
B_{n}\end{array}\right),
\end{equation}
where $M_n^D$ denotes the transfer matrix of the $n$-th barrier,
and its elements are given by
\begin{equation}
\begin{array}{ll}
\label{Element_D}
(M^D_n)_{11}=(\cos\frac{p_n a}{\hbar}+i\frac{\kappa^2+\kappa_n^2}{2\kappa\kappa_n}\sin\frac{p_n a}{\hbar})e^{-\frac{i}{\hbar}pa},\\
(M^D_n)_{12}=(i\frac{\kappa_n^2-\kappa^2}{2\kappa\kappa_n}\sin\frac{p_n a}{\hbar})e^{-\frac{i}{\hbar}p(x_n+x'_n)}, \\
(M^D_n)_{21}=(M^D_n)_{12}^*,\ \ \ \ (M^D_n)_{22}=(M^D_n)_{11}^*
\end{array}
\end{equation}
with $\kappa=(E-mc^2)/(cp)$ and $E^2=c^2p^2+m^2c^4$.

For comparison, we recall the results for non-relativistic case
which is described by the Schr\"{o}dinger equation
$[-\frac{\hbar^2}{2m}\frac{d^2}{dx^2}+V (x)-E_k]\Phi(x)=0,$ where
$E_k$ is non-relativistic kinetic energy. In this case, we have a
relation similar to Eq. (\ref{M_D}), but $M_n^D$ is replaced by
$M_n^S$ which is the transfer matrix for the $n$th barrier
calculated by the Schrodinger equation. It is straightforward to
derive that the elements of the matrix $M_n^S$ have the same form
of Eq. (\ref{Element_D}), but $\kappa$, $p_n$ (and $\kappa_n$) are
replaced by the non-relativistic counterparts $p^S=\sqrt{2mE_k}$
and $p_n^S=\sqrt{2m(E_k-V_n)}$, respectively.

We consider an incoming relativistic particle with energy $E$ ($E_k$
for non-relativistic kinetic energy ) from the left, then the
amplitude of the outgoing at the right side of the $N$ barriers is
related to the amplitude of the incoming by the relation
$\left(\begin{array}{c}
  A_{r} \\
  B_r \\
\end{array}\right)=M^J\left(\begin{array}{c}
  A_{l} \\
  B_{l}
\end{array}\right),$
where the
total transfer matrix $M^J$ (hereafter we use the superscripts
$J=D$ and $S$ to denote the relativistic and non-relativistic
cases, respectively) for the $N$ barriers reads
\begin{equation}
M^J=M^J_N D^J M^J_{N-1}\cdots D^J M^J_2D^J M^J_1.
\end{equation}
Here $D^J=\text{diag}\{ \exp(-i p^J d/\hbar), \exp(i p^J
d/\hbar)\}$  represents the displacement matrix between two
nearest neighbor barriers, with $p^D=\sqrt{E^2/c^2-m^2 c^2}$.

The transport properties can be extracted from the transfer
matrices for both relativistic and non-relativistic cases. At zero
temperature, the conductance through the $N$ barriers is given by
the Landauer formula\cite{Landauer} $G^J=\frac{e^2}{h} g^J,$ where
$g^J=1/|(M^J)_{11}|^2$ is the dimensionless conductance.
It is noted that the localization length $\xi^J$ or the Lyapunov
exponent $\gamma^J$
is defined as
$\gamma^J \equiv 1/\xi^J=-\lim_{L\rightarrow\infty} {\langle\ln
g^J\rangle}/{L}$,
where $L$ is the total length of the chain $L=Nb=N$ ( we choose
$b=d+a$ as the unit of length), and $\langle \cdots \rangle$
denotes the averaging over the disorders.  The $\xi^J$ is a
function of the energy $E$ $(E_r)$ and can be used to characterize
a localized state: a state is a localized state if $\xi^J$ is
finite and is a delocalized (extended) state if $\xi^J$ is
divergent.

It is hard to obtain an analytical expression for the localization
length $\xi^J$ 
in a general case; however, it has been
shown that   $ \lim_{L\rightarrow\infty}(\langle\ln g^S\rangle/L)$
always exists for any energy of a non-relativistic particle in an
arbitrary weak one-dimensional disordered system\cite{Lee}.  In a
similar way, we can find that $ \lim_{L\rightarrow\infty}(\langle\ln
g^D\rangle/L)$ also exists for relativistic massive particles. The
numerical procedure for both relativistic and non-relativistic cases
is as follows: one can define
$\alpha_N^J=\frac{1}{N_c}\sum_{i=1}^{N_c}\frac{1}{N}\ln g_i^J (N),$
where $g_i^J (N)$ is the conductance for a specific configuration of
fixed $N$ barriers, and then  $\alpha_N^J$ is an averaged quantity
for the number of $N_c$ configurations.
 We find that, for both non-relativistic and relativistic massive
 particles, a convergent value $\alpha_N^J$ can always be derived for sufficient large $N$, which
 could be
 considered as the localization length $\xi^J$ of the system.
 The result implies that the state for the massive Dirac particles is also a localized state
 for arbitrarily weak disorders, as in the
 non-relativistic case. The localization length (Lyapunov exponent) as a function of
 the potential width $a$
 is plotted in Fig.2. It is seen that
  the localization length of the relativistic particles is always longer than that of the
 non-relativistic particles.

We now turn to examine the validity of the single-parameter scaling
equation
  $\beta^J\equiv \frac{d\ln g^J}{d\ln L}$ in the relativistic region.
 The scaling quantities $\beta^J$ as a
function of $\langle\ln g^J \rangle$ for $m=2.5 \times 10^{-4}m_0$
with $m_0$ the mass of a nuclear in both relativistic and
non-relativistic cases are plotted in Fig.2(b). The
non-relativistic case was studied in Ref.\cite{Economou}, and our
results are essentially the same as those presented there. It is
seen clearly from Fig.2(b) that the assumption of the
single-parameter scaling theory is still valid for the massive
Dirac particles in the disordered systems.

It is interesting to note that an analytical result for a massless
particle $(m=0)$ can be derived, from which one is able to find
unexpectedly that the particle is entirely delocalized for arbitrary
energy. For massless particles, the transfer matrix of total N
barriers is obtained as $M=\text{diag}\{ e^{i\varphi/\hbar },\ \ \
e^{-i\varphi/\hbar}\}$
with $\varphi= -Npb+ \sum_{n=1}^N p_na$. The transmission amplitude
$t=\exp(i\varphi/\hbar)$
is a pure phase factor, and  the dimensionless conductance
$g^D\equiv 1$. In this case, the localization length $\xi^D$ for the
massless particles approaches to infinity,
and thus breaks down the famous conclusion that the particles are
always localized for any weak disorder in one-dimensional systems.

The inherent physics is simply
the chiral symmetry. The time-independent Hamiltonian for the Dirac
particles is given by $\label{H_D} H_D=-i\hbar
c\sigma_x\frac{d}{dx}+mc^2 \sigma_z +V(x),$ and the chiral operator
for a Dirac spinor is the matrix $\gamma^5=\sigma_x$ in
one-dimensional cases. Under the discrete chiral transformation the
spinor is transformed as $\psi_c=\gamma^5\psi$ and the transformed
Hamiltonian
\begin{equation}
\label{H_c} H_c=\gamma^5 H_D \gamma^5=-i\hbar
c\sigma_x\frac{d}{dx}-mc^2\sigma_z+V(x).
\end{equation}
Then the chirality is conserved for a massless particle.
Noting
that $\gamma^5 \phi_{\pm}=\pm \phi_{\pm} $ with $\phi_{\pm}= \left(\begin{array}{c}1\\
\pm\kappa\end{array}\right)$, the general solution of the massless
Dirac equation described in Eq.(\ref{Solution}) can be rewritten as
$\psi(x)=A \phi_{+} e^{\frac{i}{\hbar}px}+
B\phi_{-}e^{-\frac{i}{\hbar}px},$ i.e., the first (second) term is
actually the eigenstate of the chiral operator with positive
(negative) chirality. Assuming that the incoming wave function is
$\psi_{in}(x)=A \phi_{+} e^{\frac{i}{\hbar}px}$, then the outgoing
wave-function $\psi_{out}(x)=A^\prime \phi_{+}
e^{\frac{i}{\hbar}px}+ B^\prime\phi_{-}e^{-\frac{i}{\hbar}px}$
should have the same positive chirality, i.e., the reflection rate
$B^\prime$ must be zero for massless particles because of the
conservation of the chirality. However, the chirality is not
conserved for a massive particle, so the reflection $B'$ in
principle can not be always zero. In this case the massive particle
should be localized for any weak disorders.
Alternatively, an intuitive picture of
localization of relativistic particles may be understood with the
Klein's paradox. For massive particles, when the height of the
potential barrier reaches the order of $mc^2$, the barrier becomes
nearly transparent.
Since some of the barriers
are transparent, the localization length increases. For the
massless particles $mc^2$ is zero and thus every barrier is
transparent. Therefore, massless particles are always delocalized.

We now turn to address how to simulate the relativistic particles
with cold atoms.
We consider the adiabatic motion of atoms having a tripod level
configuration in the field of three laser beams, as shown in
Fig.1(b,c)\cite{Juzeliunas,Stanescu,Vaishnav}. The ground states
$|1\rangle$,$|2\rangle$ and $|3\rangle$ are coupled to an excited
state $|0\rangle$ through spatially varying laser fields, with the
corresponding Rabi frequencies $\Omega_1$,$\Omega_2$ and
$\Omega_3$, respectively. The full quantum state of the atoms
$\Phi (\mathbf{r})$
can be described as $\Phi (\mathbf{r})=\sum_{j=0}^3
\phi_j(\mathbf{r})|j\rangle$, where $\mathbf{r}$ is the atomic
position. The original Hamiltonian of the atom with the mass $m_a$
takes the form
$H=\frac{\mathbf{P}^2}{2m_a}+V_{H}(\mathbf{r})+V_{L}(\mathbf{r})+H_{int}$,
where $V_{H}(\mathbf{r})\equiv \sum_{j=0}^3
V_j^H(\mathbf{r})|j\rangle\langle j|$ represents an external
harmonic trapping potential,
and $V_{L}(\mathbf{r})$ denotes a state-independent random
potential. $H_{int}$ is the laser-atom interaction Hamiltonian
given by $ H_{int}=-\hbar\sum_{j=1}^3 (\Omega_j |0\rangle\langle
j|+\text{H.c.})$, where
the Rabi frequencies are chosen as $\Omega_1=\Omega \sin\theta e^{-i
k x}/\sqrt{2}$, $\Omega_2=\Omega \sin\theta e^{i k x}/\sqrt{2}$, and
$\Omega_3=\Omega \cos\theta e^{-i k y}$. Here
$\Omega=\sqrt{|\Omega_1|^2+|\Omega_2|^2+|\Omega_3|^2}$, $k$ is the
laser wave vector
and the angle $\theta$ defines the relative
intensity\cite{Juzeliunas,Stanescu,Vaishnav}. Following
Ref.\cite{Juzeliunas}, one is able to obtain an effective
one-dimensional Dirac-type Hamiltonian as
\begin{equation}\label{H_k} H_{k} \approx c_\star \sigma_x p_x
+\gamma_z\sigma_z+V_1^H(x)+V_L(x),
\end{equation}
up to an irrelevant constant, provided that the wave vector of the
atoms $ p_x /\hbar \ll k \cos\theta$\cite{Note_1D}. Here
 $\gamma_z\equiv \frac{\hbar^2 k^2}{2 m_a}\sin^4\theta$,
$c_\star=\frac{\hbar k}{ m_a}\cos\theta$ is the effective 'speed of
light'. In the derivation, we have assumed that the trapping
potential $V^H(\mathbf{r})$ is independent on the internal states.
Comparing the original Dirac Eq.(\ref{Hamiltonian}) with the
Dirac-like Eq.(\ref{H_k}) achieved in cold atoms, the effective
'speed of light' in cold atoms is $c_\star$ and the effective mass
$m=\frac{m_a}{2} \tan^2\theta\sin^2\theta$. Note that the mass $m$
of the simulated Dirac particle is not the mass $m_a$ of the cold
atom itself and it is a remarkable feature that the mass $m$ in
the simulated Dirac-like equation
can be controlled by the laser beams. Thus both the massive and
massless Dirac equations can be realized with cold
atoms\cite{Note}.

Finally, we  discuss briefly the detection of the localization
length. For concreteness, we assume that the tripod-level
configuration in Fig.1(b) is provided by the atoms of $^{87}$Rb,
where the ground states $\{|1\rangle, |2\rangle,|3\rangle\}$ are
the hyperfine states $5^{2}S_{1/2}(F=1,m_F=-1,0,1)$ and the
excited state $|0\rangle$ is given by either the state of
$5^{2}P_{3/2}(F=0)$ or $5^{2}P_{3/2}(F=2,m_F=0)$. In this case,
the harmonic potential $V_H(\mathbf{r})$ has been experimentally
realized by the far-off resonant laser beams in the implementation
of the spinor condensates of $^{87}$Rb\cite{Chang}. In addition, a
feasible approach to detect the localization length of the
relativistic particles can follow the case in non-relativistic
particles implemented in Ref.\cite{Billy}, except that three
additional laser beams represented by $\Omega_j$  are required.
The experiment starts with an elongated cluster of ultracold
$^{87}$Rb atoms trapped by a harmonic potential $V_H$. A
far-off-resonance laser beam (such as wavelength $1.06 \mu m$ used
in Ref.\cite{Billy}) creates an optical wave guide along the
horizonal $x$ axis, and a loose longitudinal trap is also realized
by such laser beam. Three laser beams with the resonant wavelength
of rubidium ( wavelength 0.78 $\mu m$, near the resonant also
works) but different polarizations, as shown in Fig.1(c), are
shined on the atoms to create an atomic gas that could be
described by the Dirac equation. The longitudinal confinement is
switched off at time $t=0$, and the atomic gas starts to expand in
the guide along the $x$ direction. A disordered potential $V_L$ is
applied to the expanding atomic gas using an optical speckle field
produced by passing a laser beam (with the wavelength about 0.514
$\mu m$\cite{Billy}) through a diffusing plate. One then detects
the spatial distribution of the atoms at increasing evolution time
using absorption imaging. As in the experiments\cite{Billy,Roati},
one can directly measure the localization length of the particles
by the density profiles. Therefore, the comparison of Anderson
localization between relativistic and non-relativistic cases can
be made for the conditions with and without the additional laser
beams $\Omega_j$.
Considering that the two main ingredients to observe
(de)localization of Dirac particles, the disordered potentials
\cite{Billy}  and light-induced gauge field\cite{Lin}, have been
achieved in the recent experiments on the atoms of $^{87}$Rb, it is
expected that the cold atoms may offer a novel platform for the
study of Anderson localization in the relativistic region.



In summary, we
have found that the relativistic particles tend to delocalization
and revealed that the massless  ones 
are entirely delocalized
in disordered
one-dimensional systems. The predicted features may be tested by
future experiments with ultra-cold atoms. On the other hand, the
(de)localization of the relativistic particles may also be observed
in a disordered graphene, where Dirac
electrons are confined 
to move in one
dimension and the impurities are small enough such that the
scattering does not occur between the two Dirac points.

We thank L.-M. Duan for many fruitful discussions. This work was
supported by the RGC of Hong Kong,
the NSFC (No. 10674049), and the
SKPBRC (Nos. 2006CB921800 and 2007CB925204).


\begin{figure}[tbhp]
\vspace{0.2cm}
\includegraphics[width=7cm,height=4.2cm]{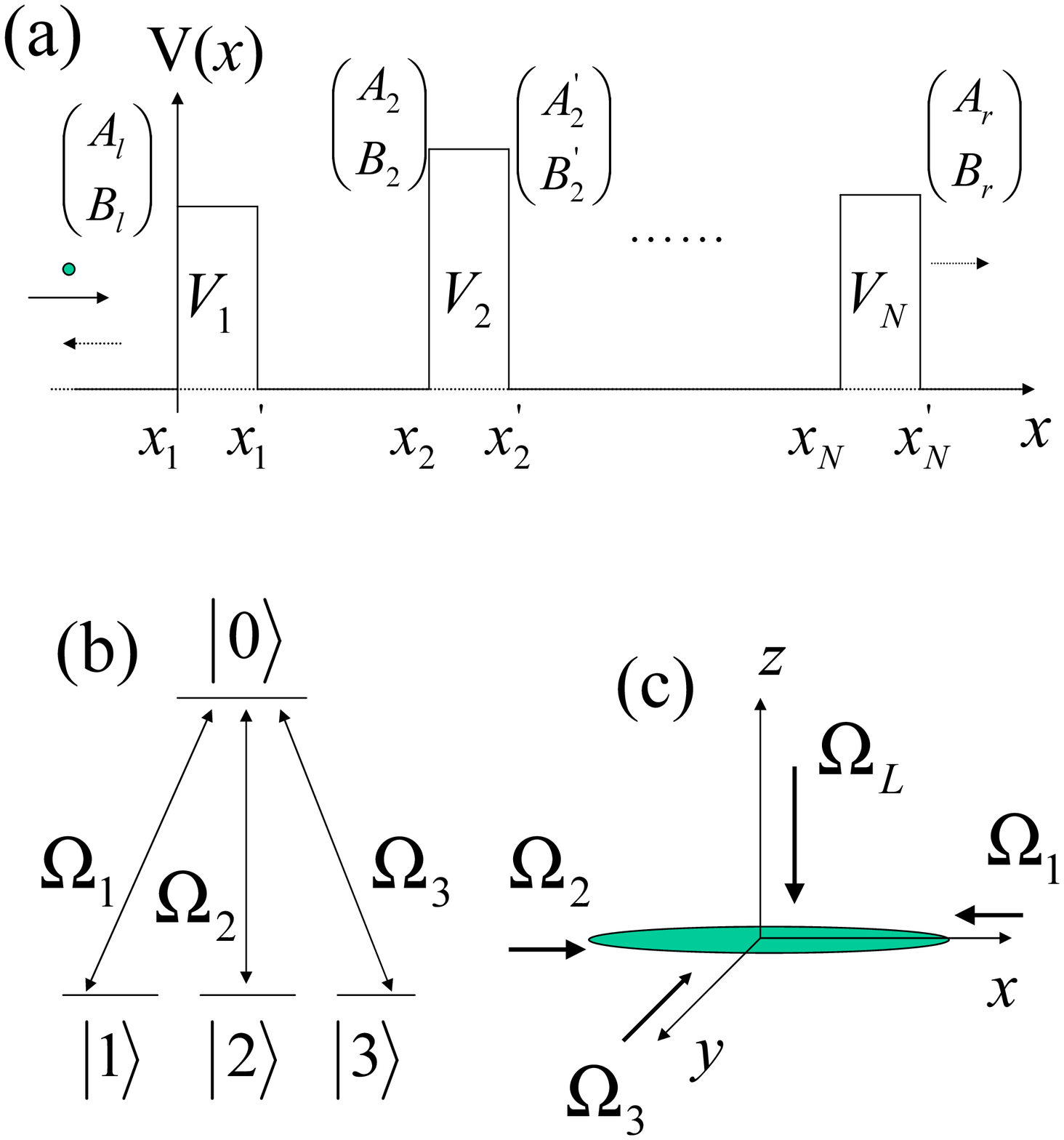}
\caption{Schematic representation of the system. (a) N rectangular
potentials.
(b) Atom with tripod level structure interacting with three laser
beams.
(c) The configuration of the laser beams to realize a Dirac-like
equation with the lasers $\Omega_j$ and a disordered potential
with the laser $\Omega_L$. The atoms are confined in a one
dimensional wave guide by a harmonic trap.}
\end{figure}

\begin{figure}[tbhp]
\includegraphics[width=7cm,height=5cm]{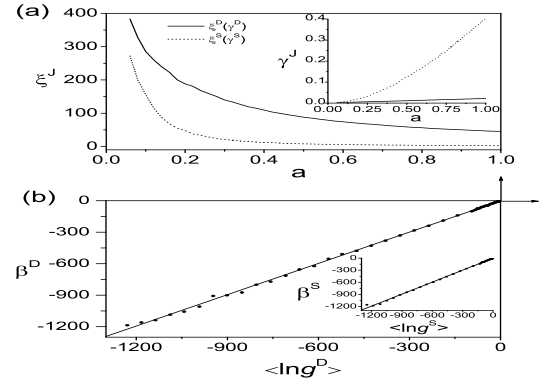}
\caption{ (a) The localization length $\xi^J$ as a function of the
potential width $a$. The insert is the corresponding Lyapunov
exponent $\gamma^J$.(b) The scaling quantity $\beta^D$ ($\beta^S$
in the insert) as a function of $\langle\ln g^D \rangle$.
$\langle
\ln g^J\rangle$ is the average of 1000 configurations. The other
parameters are $E=1.05m c^2$ and $\delta\in [-2,2]$ with the units
of $mc^2$ . }
\end{figure}


\begin{thebibliography}{10}
\bibitem{Note_1D}
We here focus on the case with an extremely anisotropic potential,
which is sufficiently strong along the transverse direction of the
$x$ axis, such that the original massive atoms are actually, or at
least in principle, confined in a one-dimensional guide along the
$x$ axis.
 A crucial condition of this Dirac equation is
$p_\alpha/\hbar \sim 1/l_\alpha \ll k\cos\theta $
$(\alpha=x,\rho)$, where $l_x$ ($l_\rho$) is the longitudinal
(transverse) confinement and $k \sim 8.0\times 10^6 m^{-1}$ for
the atoms of $^{87}$Rb. If we simply take $\theta=\arccos(1/8)$,
this condition implies that the dynamics of the atoms along the
transverse direction is nonrelativistic for $l_\rho \leq 1.0\mu m$
and thus the atoms may still be confined by the original trap,
while it is relativistic along the $x$ axis for $l_x \gg 1.0\mu
m$.
\bibitem{Note}
If we choose a specific potential given by
$V_1^H -V_3^H=\hbar^2 k^2\sin^2\theta/2m_a$, we are able to obtain
a massless one-dimensional Dirac equation as  the $\gamma_z$ in
Eq. (\ref{H_k}) vanishes.
\end{thebibliography}
\end{document}